\def\tr{\hbox{tr}}
\def\ln{\ell{n}}
\documentstyle[a4,12pt]{article}
\begin{document}
\begin{titlepage} \vspace{0.2in} \begin{flushright}
MITH-96/10 \\ \end{flushright} \vspace*{1.5cm}
\begin{center} {\LARGE \bf  A Possible Lattice Chiral Gauge Theory\\} \vspace*{0.8cm}
{\bf She-Sheng Xue$^{(a)}$}\\ \vspace*{1cm}
INFN - Section of Milan, Via Celoria 16, Milan, Italy\\ \vspace*{1.8cm}
{\bf   Abstract  \\ } \end{center} \indent
 
We analyze the dynamics of an $SU_L(2)\otimes U_R(1)$ chiral gauge theory on a
lattice with a large multifermion coupling $1\ll g_2 < \infty$. It is shown
that no spontaneous symmetry breaking occurs; the ``spectator'' fermion
$\psi_R(x)$ is a free mode; doublers are decoupled as massive Dirac fermions
consistently with the chiral gauge symmetry. Whether right-handed three-fermion
states disappear and chiral fermions emerge in the low-energy limit are
discussed. Provided right-handed three-fermion states disappear, we discuss the
chiral gauge coupling, Ward identities, the gauge anomaly and anomalous
$U_L(1)$ global current within the gauge-invariant prescription of
renormalization of the gauge perturbation theory. 

\vfill \begin{flushleft}  June, 1996   \vspace*{3cm} \\
\noindent{\rule[-.3cm]{5cm}{.02cm}} \\
\vspace*{0.2cm} \hspace*{0.5cm} ${}^{a)}$ 
E-mail address: xue@milano.infn.it\end{flushleft} \end{titlepage}
 
\noindent
{\bf 1. Introduction}
\vskip1cm

Since the ``no-go'' theorem \cite{nn81} of Nielsen and Ninomiya was
demonstrated in 1981, the problem of chiral fermion ``doubling'' and
``vector-like'' phenomenon on a lattice still exists if one insists on
preserving chiral gauge symmetries. Actually, the essential spirit of the
``no-go'' theorem of Nielsen and Ninomiya is that, under certain prerequisites,
the paradox concerning chiral gauge symmetries, vector-like doubling and
anomalies are unavoidably entwined. In fact, if taking the attitude of
retaining the chiral gauge symmetries at short distances, we regard the
``no-go'' theorem is not a real barrier, rather, its prerequisites hint what
could be the possible {\it choices of Nature} for the Standard Model at short
distances\cite{ep,xue,ns92,b92} to get resolution of this paradox. That was the
reason and motivation that we persist on preserving exact chiral gauge
symmetries by multifermion couplings, which violate bilinearity, one of the
prerequisites of the ``no-go'' theorem. 

Eichten and Preskill (EP) \cite{ep} were pioneers on the idea of
multifermion couplings ten years ago. The crucial points of their idea can be
briefly described as follows. Multifermion couplings are introduced such that,
in the phase space of strong-couplings, Weyl states composing three elementary
Weyl fermions (three-fermion states) are bound. Then, these three-fermion
states pair up with elementary Weyl fermions to be Dirac fermions. Such Dirac
fermions can be massive without violating chiral gauge symmetries due to the
appropriate quantum numbers and chirality carried by these three-fermion
states. The binding thresholds of such three-fermion states depend on
elementary Weyl modes residing in different regions of the Brillouin zone. If
one assumes that the spontaneous symmetry breaking of the Nambu-Jona Lasinio
(NJL) type \cite{njl} does not occur and such binding thresholds separate the
weak-coupling symmetric phase from the strong-coupling symmetric phase, there
are two possibilities to realize the continuum limit of chiral fermions in
phase space. One is of crossing over the binding threshold of the three-fermion
state of chiral fermions; another is of a wedge between two thresholds, where
the three-fermion state of chiral fermions has not been formed, provided all
doublers sitting in various edges of the Brillouin zone have been bound to be
massive Dirac fermions and decouple. 

To visualize this idea, EP proposed a model \cite{ep} of multifermion couplings
with $SU(5)$ and $SO(10)$ chiral symmetries and suggested the possible regions
in phase space to define the continuum limit of chiral fermions. However, the
same model of multifermion couplings with $SO(10)$ chiral symmetry was studied
in ref.~\cite{gpr}, where it was pointed out that such models of multifermion
couplings fail to give chiral fermions in the continuum limit. The reasons are
that an NJL spontaneous symmetry breaking phase separating the strong-coupling
symmetric phase from the weak-coupling symmetric phase, the right-handed Weyl
states do not completely disassociate from the left-handed chiral fermions and
the phase structure of such a model of multifermion couplings is similar to
that of the Smit-Swift (Wilson-Yukawa) model \cite{ss}, which has been very
carefully studied and shown to fail \cite{p}. 

We should not be surprised that a particular class of multifermion coupling or
corresponding Yukawa coupling models does not work. This does not means that
EP's idea is definitely wrong in all possible classes of multifermion coupling
or corresponding Yukawa coupling models, unless there is another generalized
``no-go'' theorem on interacting theories\footnote{Ref.\cite{ys93} discussed
the constraints on the existence of chiral fermions in interacting lattice
theories.}. Actually, Nielsen and Ninomiya gave an interesting comment on EP's
idea based on their intuition of anomalies \cite{nn91}. In this paper, we
present a possible lattice chiral gauge theory with an extremely large
multifermion coupling in section 2. In successive sections, we attempt to study
the problems concerning the lattice chiral gauge theory listed as fellow: 

\begin{itemize}
\begin{enumerate}
\item the ``spectator'' fermion $\psi_R$ is a free mode and decoupled; 

\item no NJL spontaneous chiral symmetry breaking occurs;

\item all doublers $p=\tilde p+\pi_A$\footnote{All momenta are scaled to
be dimensionless and the physical
momentum $\tilde p\simeq 0$ and $\pi_A$ runs over fifteen lattice momenta
$\pi_A\not=0$.} are {\it strongly bound} to be massive
Dirac fermions and decoupled consistently with the $SU_L(2)\otimes U_R(1)$
chiral symmetry;

\item the chiral fermions ($p=\tilde p$) of $\psi^i_L(x)$ and $\psi_R(x)$ have not
yet been bound to the three-fermion state, an undoubled chiral mode of
$\psi^i_L(x)$ exists in the low-energy spectrum;

\item the chiral gauge coupling and Ward identities; 

\item the vacuum functional and chiral gauge anomaly;

\item the anomalous global current (fermion number non-conservation).

\end{enumerate}
\end{itemize}
We discuss the chance that EP's idea can be realized in this model, 
and the possibilities that the model can go to failure in the end of each 
section.

\vskip1cm
\noindent
{\bf 2. Formulation and the large multifermion coupling $1\ll g_2 <\infty $}
\vskip0.7cm

Let us consider the following fermion action of the $SU_L(2)\otimes U_R(1)$ 
chiral symmetries on a lattice with one external multifermion coupling 
$g_2\gg 1$. 
\begin{eqnarray}
S&\!=\!&{1\over 2a}\sum_x\left(\bar\psi^i_L(x)\gamma_\mu D^\mu\psi^i_L
(x)+\bar\psi_R(x)\gamma_\mu\partial^\mu\psi_R(x)\right)\label{action}
\\
&\!+\!&
g_2\bar\psi^i_L(x)\!\cdot\!\Delta\psi_R(x)
\Delta\bar\psi_R(x)\!\cdot\!\psi_L^i(x),\nonumber
\end{eqnarray}
where ``$a$'' is the lattice spacing; $\psi^i_L$ ($i=1,2$) is an $SU_L(2)$
gauged doublet, $\psi_R$ is an $SU_L(2)$ singlet and both are two-component
Weyl fermions. The $\psi_R$ is treated as a ``spectator'' fermion. The $\Delta$
that is a high order differential operator on the lattice obeys 
\begin{equation}
\Delta(x)\psi_R(x)=\sum_\mu\left(\psi_R(x+\mu)+\psi_R(x-\mu)-2\psi_R(x)\right).
\label{delta}
\end{equation}
The multifermion coupling $g_2$ is a dimension-10 operator relevant only for
doublers $p=\tilde p+\pi_A$, but irrelevant for chiral fermions $p=\tilde p$ of
$\psi^i_L$ and $\psi_R$. The action (\ref{action}) preserves the global
chiral symmetry $SU_L(2)\otimes U_R(1)$ and the $\psi^i_L(x)$ can be gauged to
have the exact local $SU_L(2)$ chiral gauge symmetry. In addition, the action
(\ref{action}) possesses a $\psi_R$-shift-symmetry\cite{gp}
($\epsilon=$const.), 
\begin{equation}
\psi_R(x) \rightarrow \psi_R(x)+\epsilon.
\label{shift}
\end{equation}
The global symmetry $U_L(1)$ relating to the conservation of the fermion number
of $\psi_L^i(x)$ is explicit in eq.(\ref{action}). However, this global
symmetry will be discussed and shown to be anomalous in section 8. 

Our goal is to seek a possible segment $1\ll g_2\infty$, where an undoubled
$SU_L(2)$-chiral-gauged fermion content is exhibited in the continuum limit
consistently with $SU_L(2)\otimes U_R(1)$ chiral symmetry and the theory has
the correct features of the chiral gauge coupling, Ward identities,
gauge anomaly and anomalous global current, which are listed in introduction
section. We are bound to demonstrate these properties of the theory in this
segment. 

To prove the first and second points concerning free ``spectator'' fermion
$\psi_R(x)$ and no NJL spontaneous symmetry breaking, we need the Ward identity
stemming from the $\psi_R$-shift-symmetry (\ref{shift}). Considering the
generating functional $W(\eta,J)$ of the theory, 
\begin{eqnarray}
W(\eta,J)&=&-\ln Z(\eta,J),\hskip0.5cm 
\int_\phi=\int [d\psi_L^i d\psi_R d A_\mu]
\label{part}\\
Z(\eta,J)&=&\int_\phi\exp\left(-S+\int_x\left(\bar\psi^i_L\eta_L^i+
\bar\eta_L^i\psi^i_L+\bar\psi_R\eta_R+
\bar\eta_R\psi_R+A_\mu J^\mu\right)\right),\nonumber
\end{eqnarray}
we define the generating functional of one-particle irreducible vertices
(the effective action
$\Gamma(\psi'^i_L,\psi_R', A_\mu')$ as
the Legendre transform of $W(\eta,J)$
\begin{equation}
\Gamma(\psi'^i_L,\psi'_R,A'_\mu)=W(\eta,J)-\int_x\left(\bar\psi'^i_L\eta_L^i+
\bar\eta_L^i\psi'^i_L+\bar\psi'_R\eta_R+
\bar\eta_R\psi'_R+A'_\mu J^\mu\right),\nonumber
\end{equation}
and with the relations (for short $X=L$ and the $SU_L(2)$ index ``$i$'' 
is dropped for $X=R$)
\begin{eqnarray}
A'_\mu(x)&=&\langle A_\mu (x)\rangle=-{\delta W\over\delta J_\mu (x)},
\hskip0.3cm J_\mu(x)=-{\delta\Gamma\over\delta A'_\mu(x)},\nonumber\\
\psi'^i_X(x)&=&\langle\psi^i_X(x)\rangle =-{\delta W\over\delta\bar\eta^i_X(x)
},\hskip0.3cm
\bar\eta^i_X(x)={\delta\Gamma\over\delta\psi'^i_X(x)},\nonumber\\
\bar\psi'^i_X(x)&=&\langle\bar\psi^i_X(x)\rangle={\delta W\over\delta
\eta^i_X(x)},\hskip0.3cm 
\eta^i_X(x)=-{\delta\Gamma\over\delta\bar\psi'^i_X(x)},
\label{pl}
\end{eqnarray}
in which the fermionic derivatives are left-derivatives.
In eq.(\ref{pl}), the
$\langle\cdot\cdot\cdot\rangle$ indicates an expectation value with respect to 
the partition functional $Z(\eta,J)$ (\ref{part}).

Making the parameter $\epsilon$ in eq.(\ref{shift}) to be spacetime
dependent, and varying the generating function (\ref{part}) according to the
transformation rule (\ref{shift}) for arbitrary $\epsilon(x)\not= 0$, we arrive
at the Ward identity
corresponding to the $\psi_R$-shift-symmetry of the action (\ref{action}): 
\begin{equation}
{1\over 2a}\gamma_\mu\partial^\mu\psi^\prime_R(x)
+g_2\!\langle\Delta\!\left(\bar\psi^i_L(x)\!\cdot\!
\Delta\psi_R(x)\psi_L^i(x)\right)\rangle-{\delta\Gamma\over\delta\bar
\psi'_R(x)}=0.
\label{w}
\end{equation}
Based on this Ward identity (\ref{w}), one can get all one-particle irreducible
vertices containing at least one external ``spectator'' fermion $\psi_R$. 

Taking functional derivatives of eq.(\ref{w}) with respect to appropriate
``prime'' fields (\ref{pl}) and then putting external sources 
$J=\eta=0$, one can derive:
\begin{eqnarray}
\int_xe^{-ipx}
{\delta^{(2)}\Gamma\over\delta\psi'_R(x)\delta\bar\psi'_R(0)}\!&=&\!{i\over a}
\gamma_\mu\sin (p^\mu),\label{free}\\
\int_xe^{-ipx}
{\delta^{(2)}\Gamma\over\delta\psi'^i_L(x)\delta\bar\psi'_R(0)}\!&=&\!
{1\over2}\Sigma^i(p)\!=\!2g_2w(p)\langle\bar\psi^i_L(0)\cdot
\Delta\psi_R(0)\rangle_\circ,
\label{ws2}
\end{eqnarray}
where the $\langle\cdot\cdot\cdot\rangle_\circ$ indicates an expectation value
with respect to the partition functional $Z(J,\eta)$ (\ref{part}) with vanishing
external sources $(J=\eta=0)$. The $w(p)$ is the 
Fourier transformation of the differential operator ${1\over2}\Delta(x)$ 
(\ref{delta}),
\begin{equation}
w(p)=\sum_\mu(1-\cos p_\mu)\rightarrow O(p^2)\hskip0.5cm p\rightarrow 0,
\label{wilson}
\end{equation}
which is the Wilson factor \cite{wilson}. The expectation value $\langle\bar
\psi^i_L(0)\cdot \Delta\psi_R(0)\rangle_\circ$ can be zero (symmetric phase)
and non-zero (broken phase) depending the value of the multifermion coupling 
$g_2$ \cite{xue95}.
Other 1PI interacting vertices containing more than two external ``spectator''
fermion  $\psi_R$ identically vanish. Eq.(\ref{free}) indicates an absence of
the wave function renormalization of $\psi_R(x)$. Thus, we can conclude that
the ``spectator'' fermion $\psi_R(x)$ is free mode for vanishing eq.(\ref{ws2})
in the continuum limit $p\rightarrow 0$. This conclusion is independent
of the finite value of multifermion coupling $g_2$. 

On the other hand, in the continuum limit $p\rightarrow 0$, one can have
similar conclusion without using the Ward identity of the
$\psi_R$-shift-symmetry to exactly obtain 1PI functions (\ref{free},\ref{ws2}).
Due to the structure
of the multifermion interaction (\ref{action}), every external ``spectator''
fermion  $\psi_R$ with momentum $p$ associates with a Wilson factor $w(p)$, the
1PI interacting vertices with an external ``spectator'' fermion $\psi_R$ goes
to zero ($O(p^2)$) as $p\rightarrow 0$. From this point of view, one can see
that eqs.(\ref{free},\ref{ws2}) must be right in $p\rightarrow 0$. 

As another important consequence of the Ward identity of the shift-symmetry,
eq(\ref{ws2}) shows, for any finite values of the multifermion coupling $g_2$,
the normal modes ($p\rightarrow 0$) of $\psi_L^i(x)$ and $\psi_R$ do not
undergo NJL spontaneous symmetry breaking in the continuum limit. In equation
(\ref{ws2}), one has 
\begin{equation}
p_\mu=0,\hskip1cm \Sigma^i(p)=0. 
\label{zero} 
\end{equation} 
In ref.\cite{xue95}, we arrived at the same conclusion 
by explicit large-$N_c$ calculations. 

It is due to these properties, we can have a possibility of finding a segment
\begin{equation}
1\ll g_2<\infty
\label{segment}
\end{equation}
without NJL spontaneous symmetry breaking for normal
modes and in meantime, doublers decouple as massive Dirac fermions via the EP
mechanism that will be discussed in section 4. This gives us, as we will see in
section 5, a loophole to have chiral fermions in the continuum limit.

\vskip1cm
\noindent
{\bf 3. No NJL spontaneous symmetry breaking}
\vskip0.7cm

In previous section, we have shown the vanishing of the spontaneous symmetry
breaking for the normal mode sector (\ref{zero}) for any values of $g_2$. In
this section, as $g_2\rightarrow\infty$, we show the total absence of the
spontaneous symmetry breaking that means not only eq.(\ref{zero}) but also 
\begin{equation}
\Sigma(p)=0\hskip0.5cm p\not=0,
\label{dzero}
\end{equation}
which indicates doublers do not undergo the spontaneous symmetry breaking as
well. Note that in ref.\cite{xue95}, it was shown by large-$N_c$ calculations
that doublers acquire the NJL spontaneous symmetry breaking masses for certain
finite value of multifermion coupling $g_2$, which is inbetween the
strong-coupling symmetric phase and the weak-coupling symmetric phase. 

Adopting the technique of strong-coupling expansion in powers of ${1\over
g_2}$, we make a rescaling of the fermion fields, 
\begin{equation} 
\psi_L^i(x)\rightarrow (g_2)^{1\over4}\psi_L^i(x);\hskip1cm
\psi_R(x)\rightarrow (g_2)^{1\over4}\psi_R(x),
\label{rescale1}
\end{equation}
and rewrite the action (\ref{action}) 
and partition function (\ref{part}) in terms of new fermion fields
\begin{eqnarray}
S_f(x)&=&{1\over 2ag_2^{1\over2}}\sum^\dagger_\mu\left(\bar\psi^i_L(x)
\gamma_\mu \psi^i_L(x+\mu)+
\bar\psi_R(x)\gamma_\mu\psi_R(x+\mu)\right)\label{rfa1}\\
S_2(x)&=&\bar\psi^i_L(x)\cdot\left[\Delta\psi_R(x)\right]
\left[\Delta\bar\psi_R(x)\right]\cdot\psi_L^i(x),
\label{rs21}
\end{eqnarray}
where the gauge field is perturbatively eliminated and 
\begin{equation}
\sum^\dagger_\mu\bar\psi^i_L(x)\gamma_\mu 
\psi^i_L(x+\mu)=\sum_\mu\left(\bar\psi^i_L(x)
\gamma_\mu\psi^i_L(x+\mu)-\bar\psi^i_L(x)\gamma_\mu\psi^i_L(x-\mu)\right).
\label{dshift}
\end{equation}
For the coupling $g_2\rightarrow\infty$, the kinetic
terms $S_f$(\ref{rfa1}) can be dropped and we consider the strong-coupling 
limit. With $S_2(x)$ given in eq.(\ref{rs21}), the integral of $e^{-S_2(x)}$ is 
given
\begin{eqnarray} 
Z&=&\Pi_{xi\alpha}\int[d\bar\psi_R^\alpha (x) d\psi_R^\alpha (x)]
[d\bar\psi_L^{i\alpha}(x) d\psi_L^{i\alpha}(x)]\exp\left(-S_2(x)\right)
\nonumber\\
&=&2^{4N}\left(\det\Delta^2(x)\right)^4,
\label{stronglimit1}
\end{eqnarray}
where the determint is taken only over the lattice-space-time and ``$N$'' is
the number of lattice sites. 
For the non-zero eigenvalues of the operator
$\Delta^2(x)$, eq.~(\ref{stronglimit1}) shows an existence of the sensible 
strong-coupling limit. However, as for the zero eigenvalue of the operator
$\Delta^2(x)$, this strong-coupling limit should not be analytic and the 
strong-coupling expansion in powers of ${1\over g_2}$ breaks down. 

To show eq.(\ref{dzero}), i.e.~ no NJL symmetry breaking occurs in this 
segment (\ref{segment}), we need to calculate the two-point functions: 
\begin{eqnarray}
S^j_{RL}(x)&\equiv&\langle\psi_R(0),\bar\psi^j_L(x)\rangle,\label{srl}\\
S^j_{RR}(x)&\equiv&\langle\psi_R(0),[\bar\psi^j_L(x)\cdot\psi_R(x)]
\bar\psi_R(x)\rangle.
\label{smr}
\end{eqnarray}
Using the strong-coupling expansion in $O({1\over g_2})$ and at non-trivial
leading order, we get recursion relations \cite{xue95}:
\begin{eqnarray}
S^j_{RL}(x)&=&{1\over g_2\Delta^2(x)}\left({1\over 2a
}\right)^3\sum^\dagger_\mu S^j_{RR}(x+\mu)\gamma_\mu,\label{re5}\\
S^j_{RR}(x)&=&{1\over g_2\Delta^2(x)}\left({1\over 2a
}\right)\sum^\dagger_\mu S^j_{RL}(x+\mu)\gamma_\mu.
\label{re6}
\end{eqnarray}
These recursion relations are not valid where the operator $\Delta^2(x)$ has zero
eigenvalue. For $p\not=0$ and $\Delta^2(p)\not=0$, the Fourier transformation of
these recursion relations leads to 
\begin{eqnarray}
S^j_{RL}(p)&=&{1\over 4g_2w^2(p)}\left({i\over 4a^3
}\right)\sum_\mu \sin p^\mu S^j_{RR}(p)\gamma_\mu,\label{rep5}\\
S^j_{RR}(p)&=&{i\over 4g_2w^2(p)a}\sum_\mu\sin p^\mu S^j_{RL}(p)\gamma_\mu,
\label{rep6}
\end{eqnarray}
where $S^j_{RR}(p)$ and $S^j_{RL}(p)$ are the Fourier transformation of 
eqs.(\ref{srl},\ref{smr}). The solution to these recursion relations is 
\begin{equation}
\left((8ag_2w^2(p))^2+{1\over a^2}\sum_\mu\sin^2p_\mu\right)S_{RL}^j(p)=0.
\label{sigmas}
\end{equation}
Clearly, for $p\not=0$, we must have
\begin{equation}
\Sigma^j(p)\sim S^j_{RL}(p)=0,\hskip0.5cm
S^j_{RR}(p)=0.
\end{equation}
This demonstration can be straightforwardly generalized to show the vanishing of
all n-point functions that are not $SU_L(2)\otimes U_R(1)$ chiral symmetric.
Together with eq.(\ref{zero}), we conclude the segment $1\ll g_2<\infty$
is entirely symmetric and no NJL spontaneous symmetry breaking takes place. 

\vskip1cm
\noindent
{\bf 4. Three-fermion states and vector-like spectrum}
\vskip0.7cm

We turn to the third point that concerns about decoupling of doublers. On
this extreme strong coupling symmetric segment (\ref{segment}), the $\psi^i_L$ and
$\psi_R$ in (\ref{action}) are bound up to form three-fermion
states\cite{ep,xue95}: 
\begin{equation}
\Psi_R^i={1\over
2a}(\bar\psi_R\cdot\psi^i_L)\psi_R;\hskip1cm\Psi^n_L={1\over 2a}(\bar\psi_L^i
\cdot\psi_R)\psi_L^i.
\label{bound}
\end{equation}
These bound states are Weyl
fermions and respectively pair up with the $\bar\psi_R$ and $\bar\psi_L^i$
to be massive, neutral $\Psi_n$ and charged $\Psi_c^i$
Dirac fermions, 
\begin{equation}
\Psi^i_c=(\psi_L^i, \Psi^i_R),\hskip1cm\Psi_n=(\Psi_L^n, \psi_R).
\label{di}
\end{equation}
These three-fermion states (\ref{bound}) carry the appropriate quantum
numbers of the chiral gauge group that accommodates the $\psi^i_L$ and
$\psi_R$. The $\Psi_R^i$ is $SU_L(2)$-covariant and $U_R(1)$ invariant. The
$\Psi_L^n$ is $SU_L(2)$-invariant and $U_R(1)$-covariant. Thus, the spectrum of
the massive composite Dirac fermions $\Psi^i_c$ and $\Psi_n$ (\ref{di}) is vector-like,
consistently with the $SU_L(2)\otimes U_R(1)$ chiral symmetry. 

In order to study 1PI vertex functions containing the external legs of 
three-fermion states (\ref{bound}), we define: (i) the
right-handed composite ``primed'' field as 
\begin{equation}
\Psi'^i_R\equiv \langle\Psi^i_R\rangle
= {1\over 2a}{\delta^{(3)}
W(\eta)\over\delta\eta_R(x)\delta\bar\eta^i_L(x)\delta\bar\eta_R(x)};
\label{com1}
\end{equation}
and (ii) the
left-handed composite ``primed'' field as 
\begin{equation}
\Psi'^n_L\equiv \langle\Psi^n_L\rangle={1\over 2a}{\delta^{(3)}
W(\eta)\over\delta\eta^i_L(x)\delta\bar\eta_R(x)\delta\bar\eta^i_L(x)}.
\label{com2}
\end{equation}
Thus, 1PI vertex functions
containing the external legs of three-fermion states (\ref{bound}),
\begin{equation}
{\delta^{(2)}\Gamma\over\delta\Psi'^i_R(x)\bar\psi'^j_L(y)};\hskip1cm
{\delta^{(2)}\Gamma\over\delta\Psi'^n_L(x)\bar\psi'_R(y)},
\cdot\cdot\cdot,
\label{1pi}
\end{equation} 
are the truncation of the Green functions 
\begin{eqnarray}
\langle\Psi^i_R(x)\bar\psi^j_L(0)\rangle &=&{1\over 2a}{\delta^{(4)}
W(\eta)\over\delta\eta_R(x)\delta\bar\eta^i_L(x)\delta\bar\eta_R(x)\delta
\eta^j_L(0)},\nonumber\\
\langle\Psi^n_L(x)\bar\psi_R(0)\rangle &=&{1\over 2a} {\delta^{(4)}
W(\eta)\over\delta\eta^i_L(x)\delta\bar\eta_R(x)\delta\bar\eta^i_L(x)\delta
\eta_R(0)},\cdot\cdot\cdot.
\label{green}
\end{eqnarray}

Taking functional derivative of the Ward identity eq.(\ref{w}) with respect to
$\Psi'^n_L(x)$ and then putting external sources $J=\eta=0$, we can derive 
\begin{equation}
\int_xe^{-ipx}
{\delta^{(2)}\Gamma\over\delta\Psi'^n_L(x)\delta\bar\psi'_R(0)}=aM(p),
\label{dis}
\end{equation}
where 
\begin{equation}
M(p)=8ag_2w^2(p).
\label{epmass}
\end{equation}
In the basis of the 1PI vertex functions eqs.(\ref{free},\ref{dis}), we can 
determine the inverse propagator of the neutral composite
Dirac fermion $\Psi_n(x)$ to be,
\begin{equation}
S_n^{-1}(p)=\sum_\mu\gamma_\mu f^\mu(p)P_L
+{i\over a}\sum_\mu\gamma_\mu \sin p^\mu P_R+M(p),
\label{sn}
\end{equation}
where $f_\mu(p)$ remains unknown.

To obtain the inverse propagator of the charged Dirac fermion $\Psi_c^i(x)$ and $f_\mu(p)$ in
eq.(\ref{sn}), we have to use the strong-coupling expansion in powers of
${1\over g_2}$ to compute the following two-point functions with 
insertions of appropriate composite operators, 
\begin{eqnarray}
S^{ij}_{LL}(x)&=&
\langle\psi^i_L(0)\bar\psi^j_L(x)\rangle,\hskip1cm 
S^{ij}_{RL}(x)=\langle\psi^i_L(0)\bar\Psi^j_R(x)\rangle\nonumber\\
S^{ij}_{RR}(x)&=&\langle\Psi^i_R(0)\bar\Psi^j_R(x)\rangle,\hskip1cm
S^{ij}_{LR}(x)=\langle\Psi^i_L(0)\bar\psi^j_R(x)\rangle.
\label{twopoint}
\end{eqnarray}
In the lowest nontrivial order, we obtain following recursion relations 
\cite{xue95}, 
\begin{eqnarray} 
S^{ij}_{LL}(x)&=&{1\over g_2\Delta^2(x)}\left({1\over 2a
}\right)^2\sum^\dagger_\mu S^{ij}_{RL}(x+\mu)\gamma_\mu,\label{re11}\\
S^{ij}_{RL}(x)&=&\left({1\over 2a
}\right)\left({\delta(x)\delta_{ij}\over 2g_2\Delta^2(x)}
+{1\over g_2\Delta^2(x)}\left({1\over 2a
}\right)\sum^\dagger_\mu S^{ij}_{LL}(x+\mu)\gamma_\mu\right),
\label{re21}\\
S^{ij}_{RR}(x)&=&\left({1\over 2a
}\right)^2{1\over g_2\Delta^2(x)}\sum^\dagger_\mu 
\gamma_\mu S^{ij}_{RL}(x+\mu).
\label{re31}
\end{eqnarray}
It is suggested by eq.(\ref{re21}) that the states coupling to operators
$\psi^i_L(x)$ and $\Psi^i_R(x)$ are mixed in the lowest nontrivial order of 
the expansion in ${1\over g_2}$, producing a massive four-component Dirac 
fermion. To find masses, we make the Fourier transformation of these recursion
equations for $p\not=0$ and $\Delta^2(p)=4w^2(p)\not=0$,
\begin{eqnarray}
S^{ij}_{LL}(p)&=&{1\over 4g_2w^2(p)}\left({i\over 2a^2
}\right)\sum_\mu \sin p^\mu S^{ij}_{RL}(p)\gamma_\mu,\label{rep11}\\
S^{ij}_{RL}(p)&=&\left({1\over 2a
}\right)\left[{\delta_{ij}\over 8g_2w^2(p)}
+{i\over 4g_2w^2(p)a}\sum_\mu\sin p^\mu S^{ij}_{LL}(p)\gamma_\mu\right],
\label{rep21}\\
S^{ij}_{RR}(p)&=&\left({1\over 2a
}\right)\left[{1\over 4g_2w^2(p)}\left({i\over a
}\right)\sum_\mu \sin p^\mu \gamma_\mu S^{ij}_{RL}(p)
\right],
\label{rep31}\\
\end{eqnarray}
and obtain
\begin{eqnarray}
S^{ij}_{LL}(p)&=&P_L{\delta_{ij}{i\over a}\sum_\mu\sin p^\mu\gamma_\mu\over
{1\over a^2}\sum_\mu\sin^2 p_\mu+M^2(p)}P_R,\label{sll21}\\
S^{ij}_{RL}(p)&=&P_L{\delta_{ij}M(p)\over
{1\over a^2}\sum_\mu\sin^2 p_\mu+M^2(p)}P_L,\label{slm21}\\
S^{ij}_{LR}(p)&=&P_R{\delta_{ij}M(p)\over
{1\over a^2}\sum_\mu\sin^2 p_\mu+M^2(p)}P_R,\label{slm21'}\\
S^{ij}_{RR}(p)&=&P_R{\delta_{ij}{i\over a}\sum_\mu
\sin p^\mu\gamma_\mu\over
{1\over a^2}\sum_\mu\sin^2 p_\mu+M^2(p)}P_L.\label{smm21}
\end{eqnarray}
Finally we see that the field $\psi^i_L(x)$ couples to $\Psi_R^i(x)$ to form
Dirac massive states with masses $M(p)$ (\ref{epmass}) given by the
locations of the poles of eqs.(\ref{sll21}-\ref{smm21}). 
From these equations, we obtain the propagator of a massive Dirac fermion
\begin{eqnarray}
S^{ij}_c(p)&=&S^{ij}_{LL}(p)+S^{ij}_{RL}(p)+S^{ij}_{LR}(p)+S^{ij}_{RR}(p)
\nonumber\\
&=&\delta_{ij}{{i\over a}\sum_\mu\sin p^\mu\gamma_\mu+M(p)\over
{1\over a^2}\sum_\mu\sin^2 p_\mu+M^2(p)}\hskip1.5cm p\not=0,
\label{sdp}\\
S_c^{-1}(p)_{ij}&=&\delta_{ij}\left({i\over a}\sum_\mu\gamma_\mu \sin p^\mu P_L
+{i\over a}\sum_\mu\gamma_\mu \sin p^\mu P_R+M(p)\right).
\label{sc1}
\end{eqnarray}
Similar result can be obtained for
the neutral composite Dirac fermion (\ref{sn}) with 
\begin{equation}
f_\mu(p)={i\over a}\sum_\mu\gamma_\mu
\sin p^\mu,
\end{equation}
and the propagator of the neutral Dirac fermion is,
\begin{eqnarray}
S_n(p)&=&{{i\over a}\sum_\mu\sin p^\mu\gamma_\mu+M(p)\over
{1\over a^2}\sum_\mu\sin^2 p_\mu+M^2(p)}\hskip1.5cm p\not=0,
\label{snp}\\
S_n^{-1}(p)&=&\left({i\over a}\sum_\mu\gamma_\mu \sin p^\mu P_L
+{i\over a}\sum_\mu\gamma_\mu \sin p^\mu P_R+M(p)\right).
\label{sn1}
\end{eqnarray}
Eqs.(\ref{sdp}) and (\ref{snp}) show that all $SU_L(2)$ charged and neutral
doublers $p=\tilde p+\pi_A$ are decoupled as massive Dirac fermions
consistently with the $SU_L(2)\otimes U_R(1)$ chiral symmetries. 

These three-fermion states are
constituted by a soft fermionic pair\footnote{The momentum of this fermionic
pair is very small $q=p'-p\ll 1$, where $p'$ and $p$ are the momenta of
constituent fermions.} $\bar\psi^i_L\cdot\psi_R$ together with chiral fermions
$\psi_L^i(x)$ or $\psi_R(x)$. The structure of the multifermion interaction
(\ref{action}) shows, for each external ``spectator'' fermion $\psi_R$, there is
a Wilson factor $w(p)$ that is not vanishing for $p\sim \pi_A$. It can be shown
that, in order to have an enough strong interacting strength to bind up such
soft fermionic pair and then three-fermion states with momentum $p \simeq
\pi_A$, these three constituent fermions must have roughly equal momenta
\begin{equation}
(p,-p,p)
\label{emom}
\end{equation}
$|p|\sim\pi_A$ modulo $2\pi$. With such bound states mixing with elementary
states, it is possible to just fill the Dirac sea for single fermion modes
using the bound constituents, one could for low energy physics purposes get rid
of doublers \cite{nn91}. It is also pointed out in ref.\cite{nn91}, the
configuration (\ref{emom}) is hard to be localized because of the Heisenberg
uncertainty principal. We really should consider a superposition of different
momenta. This makes it very hard to fill some constituent momentum states up
precisely with bound constituents. We leave this discussion open for future
work to make sure the decoupling of doublers. 
 
\vskip1cm
\noindent
{\bf 5. Chiral fermions in the low-energy region}
\vskip0.7cm

In this section, we discuss the fourth point that the normal modes $(p=\tilde
p\sim 0)$ of the $\psi^i_L(x)$ and $\psi_R(x)$ are massless and chiral in the
low-energy limit. This is most difficult point to show for the time being,
since the strong coupling expansion in powers of ${1\over g_2}$ breaks down
for $p\rightarrow 0$. In
the basis of the continuity \cite{nn81,ys93} of the spectrum
(\ref{sdp},\ref{snp}) in the momentum space due to the locality of the theory
(\ref{action}), one may argue that the vector-like spectrum (\ref{sdp},\ref{snp}), which is
obtained in $p\not=0$, can be continously extrapolated on to $p\rightarrow 0$,
and we fail to have chiral fermions in the low-energy region. 

However, we would like to look at this point based on the point of view that is
the essential idea presented in the original paper of Eichten and Preskill
\cite{ep} to have chiral fermions in continuum limit. The question of whether
the spectra of normal modes ($p\rightarrow 0$) of $\psi^i_L$ and $\psi_R$ are
chiral is crucially related to the question of whether the normal modes ($p\sim
0$) of the three-fermion states (\ref{di}) have been composed in the segment
$1\ll g_2<\infty$. The effective multifermion coupling for these normal modes
becomes small and the binding energy of these three-fermion states becomes
small as $p\rightarrow 0$. The continuity of the spectrum (\ref{sdp},\ref{snp})
in the momentum space breaks down when the spectrum meets a threshold, if there
exists a such threshold in $p\rightarrow 0$, where the binding energy of these
three-fermion states goes to zero. In the following discussion, we adopt the
1+1 dimension case to illustrate this threshold phenomenon. 
 
We take the charged Dirac fermion (\ref{sdp}) on its mass shell and
consider that the time direction is continuous and one space is discrete. We
obtain the dispersion relation corresponding to this Dirac fermion
(\ref{sdp}) for $p\not=0$, 
\begin{equation}
E(p)=\pm\sqrt{\sin^2p+(8a^2gw^2(p))^2},
\label{des}
\end{equation}
where $E(p)$ is the dimensionless energy of the state ``$p$''. In
eq.(\ref{des}), the ``+'' sign corresponds to the dispersion relation of the
right-handed three-fermion state $\Psi_R^i(x)$. Due to the locality of the
theory, the spectrum of this bound state $\Psi^i_R(x)$ is continuous in the
momentum space \cite{nn81,ys93}. The vector-like spectrum (\ref{des}) that we
obtained by the strong coupling expansion at $p\sim$ O(1), can be analytically
continued to low momentum states $p\rightarrow 0$, unless this bound state
$\Psi^i_R$ hits the energy threshold of disappearing into its constituents, if
there exists a such threshold. Note that eq.(\ref{des}) obtained from the
strong coupling expansion at $p=0$ is not analytic (\ref{stronglimit1}). 

For a given total momentum $p$ in the low-energy region, we consider a system
that is not a bound state and contains the same constituents as the bound state
$\Psi_R^i(x)$ does, i.e., three free chiral fermions: right-handed fermions
$\bar\psi_R$ and $\psi_R$ with momenta $p_1$ and $p_2$; a left-handed fermion
$\psi_L^i$ with momentum $p_3$, where
\begin{equation}
p=p_1+p_2+p_3>0,\hskip0.3cm |p_i|\ll {\pi\over2},
\hskip0.3cm i=1,2,3.\label{totalm}
\end{equation}
As we have shown that the NJL spontaneous symmetry breaking does not occur 
(\ref{zero}) for the states $|p_i|\rightarrow 0$ in the segment
$(1\ll g_2<\infty)$, the total energy of such a system is given by
\begin{eqnarray}
E_t&=&E_1(p_1)+E_2(p_2)+E_3(p_3)\nonumber\\
E_1(p_1)&=& \sqrt{\sin^2 p_1},\hskip0.5cm p_1>0\nonumber\\ 
E_2(p_2)&=& \sqrt{\sin^2 p_2},\hskip0.5cm p_2>0\nonumber\\ 
E_3(p_3)&=&-\sqrt{\sin^2 p_3},\hskip0.5cm p_3<0 
\label{totale}
\end{eqnarray}
where all negative-energy states have been filled. There is no any definite
relationship between the total energy $E_t$ and the total momentum $p$, since
this system is not a bound state. The total energy $E_t$ level of such a system
is continuous because of relative degrees of freedom $(p_1,p_2,p_3)$ within the
system. 

The lowest energy $min E_t$ (the threshold) of such a system and
corresponding configuration can be determined by minimizing the following total
energy with a legendre multiple $\lambda$ (the constraint (\ref{totalm})), 
\begin{equation}
E_t=E_1(p_1)+E_2(p_2)+E_3(p_3)+\lambda p.
\label{legendre}
\end{equation}
One obtains the threshold of this system and the corresponding configuration
in the momentum space
\begin{eqnarray}
min E_t(p)&=&3|\sin p|,\label{thre}\\
p_3&=&-p,\hskip1cm p_1=p_2=p.\nonumber 
\end{eqnarray}
Note that this configuration is the same as that (\ref{emom}) of the 
three-fermion state that we discussed in the end of previous section.

On the other hand, the three-fermion state 
$\Psi^i_R\sim(\bar\psi_R\cdot\psi_L^i)\psi_R$, as a
composite particle, has the definite relationship between its energy and
momentum that is given by the dispersion relation (\ref{des}) with the ``+''
sign. Given the same momentum ``$p$''as eq.(\ref{totalm}), this composite
fermion states is stable, only if only there is an energy gap $\delta(p)$
(binding energy) between the threshold (\ref{thre}) and the energy (\ref{des})
of the three-fermion state, i.e., 
\begin{equation}
\delta(p)=min E_t(p)-E(p)>0.
\label{sta}
\end{equation}
The three-fermion state disappears into its constituents, when the energy gap 
$\delta$ goes to zero,
\begin{equation}
\delta(p)=min E_t(p)-E(p)=0.
\label{con}
\end{equation}
The same discussions can be applied to the neutral three-fermion state
$\Psi_L^n$ (\ref{snp}). This discussion is very much like the case of hydrogen,
a bound state composed by an electron and a proton, where the energy gap
between the first energy-level (n=1) and the continuous spectrum is 13.6 eV.
Hydrogen turns into a free electron and a free proton as the energy-gap
disappears ($n\gg 1$). 

Substituting eqs.(\ref{des}) and (\ref{thre}) into eq.(\ref{con}), we obtain
in the continuum limit $p\rightarrow 0$, the energy-gap
\begin{equation}
\delta(p)\rightarrow 0,\hskip0.5cm p\rightarrow 0
\label{fin}
\end{equation}
where the three-fermion-state spectrum dissolves into free chiral fermion
spectra. Obviously, this plausible speculation needs receiving either a
rigorously analytical proof or a numerical evidence, which are subject to
future work. Nevertheless, We assume there exist a threshold in momentum space.
The low-energy fermion states ``$p$'' below this threshold $\epsilon$ 
\begin{equation}
|p|<\epsilon\ll {\pi\over2},
\label{pthre}
\end{equation}
are massless and chiral. This threshold $\epsilon$ certainly depends on the
multifermion coupling $g_2$. It is right now not clear to us how to determine
this threshold. 

To end these discussions, we would like to point out the fact that the normal
modes do not undergo the NJL spontaneous symmetry breaking (\ref{zero}) for any
finite value of the multifermion coupling $g_2$ is extremely crucial. This
means, respect to normal modes, there is no a broken phase separating the
strong symmetric phase from the weak symmetric phase. In the other words, there
is no a mass-gap in eq.(\ref{totale}). Otherwise, the system (\ref{totale})
would be massive, the energy gap (\ref{sta}) would never be zero and we end up
with vector-like spectrum in the low-energy $p\rightarrow 0$ region. This is
the main reason for the failure of EP's approach, as pointed out in
ref.\cite{gpr}. Thus, we might have a chance to realize EP's idea that (i) the
chiral continuum limit can be defined on a phase transition from one symmetric
phase to another symmetric phase; (ii) there is a region in the phase space
$g_2$ where doublers are gauge-invariantly decoupled and normal modes are
chiral (non NJL-generated masses). The later is a possible case (the segment
$1\ll g_2<\infty$) that we have discovered. 

If what we expect can be convincingly confirmed, in the continuum limit,
undoubled low-energy chiral fermions $\psi^i_L(x)$ and $\psi_R(x)$ exist
consistently with the $SU_L(2)\otimes U_R(1)$ symmetry ($\tilde p$ is the
dimensionful momentum), 
\begin{equation} 
S^{-1}_L(\tilde p)^{ij}=i\gamma_\mu\tilde p^\mu\tilde Z_2\delta_{ij}P_L; 
\hskip0.5cm
S^{-1}_R(\tilde p)=i\gamma_\mu\tilde p^\mu P_R, 
\label{sf} 
\end{equation}
where $\tilde Z_2$ is the finite wave-function renormalization constant of the
elementary interpolating field $\psi_L^i(x)$\cite{xue95}. The spectrum of the
theory in this segment is the following. It consists of 15 copies of
$SU(2)$-QCD charged Dirac doublers eq.(\ref{sdp}) and 15 copies of $SU(2)$
neutral Dirac doublers eq.(\ref{snp}). They are very massive and decoupled.
Beside, the low energy spectrum contain the massless normal modes
eqs.(\ref{sf}) for $p=\tilde p$. This is very analogous to the mirror fermion
model\cite{mont}, except doubler masses are not generated by the spontaneous
symmetry breaking. 

\vskip1cm
\noindent
{\bf 6. Gauge coupling vertices and Ward identities}
\vskip0.7cm

Whether this chiral continuum theory in the scaling region $1\ll g_2<\infty$
could be altered, as the $SU_L(2)$ chiral gauge coupling $g$ is perturbatively
turned on and the action (1) is $SU_L(2)$-chirally gauged. One should expect a
slight change of scaling segment. We should be able to re-tune the multifermion
couplings $g_2$ to compensate these perturbative changes, due the fact that the
$SU_L(2)$-chiral gauge interaction does not spoil the $\psi_R$-shift-symmetry.
As a consequence, the Ward identity associating with the
$\psi_R$-shift-symmetry remains valid when the chiral gauge interaction reacts.

Based on the Ward identity of the $\psi_R$-shift-symmetry (\ref{w}), we take
functional derivatives with respect of the gauge field $A'_\mu$, and we arrive
at the following Ward identities, 
\begin{equation}
{\delta^{(2)} \Gamma\over\delta
A'_\mu\delta\bar\psi'_R}={\delta^{(3)} \Gamma\over\delta
A'_\mu\delta\psi'_R\delta\bar\psi'_R}={\delta^{(3)} \Gamma\over\delta
A'_\mu\delta\Psi'^n_L\delta\bar\psi'_R}=\cdot\cdot\cdot=0.
\label{wa}
\end{equation}
As a result of these Ward identities and identical vanishing of 1PI functions
containing external gauge fields, ``spectator'' fermions $\psi_R(x)$ and neutral
composite field $\Psi^n_L(x)$, we find that absolute non interacting
between the gauge field and the ``spectator'' fermion $\psi_R$ and the
neutral three-fermion states $\Psi_L^n(x)$. Thus, we disregard those
neutral modes.

In order to find the interacting vertex between the gauge 
boson and the charged Dirac fermion $\Psi^i_c(x)$, we need to consider the 
following three-point functions,
\begin{eqnarray}
\langle\Psi_c(x_1)\bar\Psi_c(x) A_\nu^a(y)\rangle
&=&\langle\psi_L(x_1)\bar\psi_L(x) A_\nu^a(y)\rangle
+\langle\psi_L(x_1)\bar\Psi_R(x) A_\nu^a(y)\rangle
\nonumber\\
&+&\langle\Psi_R(x_1)\bar\psi_L(x) A_\nu^a(y)\rangle
+\langle\Psi_R(x_1)\bar\Psi_R(x) A_\nu^a(y)\rangle,
\label{3points}
\end{eqnarray}
where we omit henceforth the $SU_L(2)$ indices $i$ and $j$.
Assuming the vertex functions to be $\Lambda_\mu^a(p,p')$ and $q=p'+p$,
we can write the three-point functions in the momentum space:
\begin{eqnarray}
\int_{x_1xy}e^{i(p'x\!+\!px_1\!-\!qy)}\langle\psi_L(x_1)\bar\psi_L(x) A_\nu^a(y)\rangle
\!&=&\! G^{ab}_{\nu\mu}(q)S_{LL}(p) \Lambda^b_{\mu LL}(p,p')S_{LL}(p');
\label{rpa1}\\
\int_{x_1xy}e^{i(p'x\!+\!px_1\!-\!qy)}\langle\psi_L(x_1)\bar\Psi_R(x) A_\nu^a(y)\rangle
\!&=&\! G^{ab}_{\nu\mu}(q)S_{LL}(p) \Lambda^b_{\mu LR}(p,p')S_{RR}(p');
\label{rpa2}\\
\int_{x_1xy}e^{i(p'x\!+\!px_1\!-\!qy)}\langle\Psi_R(x_1)\bar\Psi_R(x) A_\nu^a(y)\rangle
\!&=&\! G^{ab}_{\nu\mu}(q)S_{RR}(p) \Lambda^b_{\mu RR}(p,p')S_{RR}(p'),
\label{rpa3}\\
\int_{x_1xy}e^{i(p'x\!+\!px_1\!-\!qy)}\langle\Psi_c(x_1)\bar\Psi_c(x) A_\nu^a(y)\rangle
\!&=&\! G^{ab}_{\nu\mu}(q)S_c(p) \Lambda^b_{\mu c}(p,p')S_c(p'),
\label{rpad}
\end{eqnarray}
where $G^{ab}_{\nu\mu}(q)$ is the propagator of the gauge boson; the 
$S_{LL}(p), S_{RR}(p)$ and $S_c(p)$ are the propagators of chiral fermions 
$\psi_L(x), \Psi_R(x)$ and Dirac fermion $\Psi_c(x)$ given in 
eqs.(\ref{sll21}-\ref{sdp}).

Using the small gauge coupling expansion, one can directly calculate the 
three-point function
\begin{eqnarray}
&&\langle\psi_L(x_1)\bar\psi_L(x) A^a_\mu(y)\rangle=i{g\over 2}\left(
{\tau^a\over 2}\right)\sum_z\langle\psi_L(x_1)\bar\psi_L(x)\rangle\gamma_
\rho\nonumber\\
&&\left[\langle\psi_L(z\!+\!\rho)\bar\psi_L(x)\rangle\langle
A^b_\rho(z\!+\!{\rho\over2})A^a_\mu(y)\rangle\!+\!
\langle\psi_L(z\!-\!\rho)\bar\psi_L(x)\rangle\langle
A^b_\rho(z\!-\!{\rho\over2})A^a_\mu(y)\rangle\right],
\label{3p}
\end{eqnarray}
and obtains 
\begin{eqnarray}
\Lambda_{\mu LL}^{(1)}(p,p') &=& ig\left(\tau^a\over 2\right)\cos\left({p+p'\over 2}
\right)_\mu\gamma_\mu P_L,\label{normal}\\
\Lambda_{\mu\nu LL}^{(2)}(p,p') &=& -i{g^2\over 2}\left(\tau^a\tau^b\over 4\right)
\sin\left({p+p'\over 2}\right)_\mu
\gamma_\mu \delta_{\mu\nu}P_L,\nonumber\\
&\cdot\cdot\cdot.&\nonumber
\end{eqnarray}

By the strong coupling expansion in powers of ${1\over g_2}$, we try to compute
the other three-point functions in eqs.(\ref{3points}) in terms of
$\langle\psi_L(x_1)\bar\psi_L(x) A_\nu^a(y)\rangle$. Analogously to recursion
relations (\ref{re11}-\ref{re31}), we obtain the following recursion relations
at the nontrivial order, 
\begin{eqnarray}
\langle\psi_L(x_1)\bar\psi_L(x) A_\nu^a(y)\rangle
&=&{1\over g_2\Delta^2(x)}\left({1\over 2a}\right)^2\sum^\dagger_\rho
\langle\psi_L(x_1)\bar\Psi_R(x+\rho) A_\nu^a(y)\rangle\gamma_\rho
\label{ra1}\\
\langle\psi_L(x_1)\bar\psi_L(x) A_\nu^a(y)\rangle
&=&{1\over g_2\Delta^2(x_1)}\left({1\over 2a}\right)^2\sum^\dagger_\rho
\gamma_\rho\langle\Psi_R(x_1+\rho)\bar\psi_L(x) A_\nu^a(y)\rangle
\label{ra1'}\\
\langle\Psi_R(x_1)\bar\Psi_R(x) A_\nu^a(y)\rangle
&=&{1\over g_2\Delta^2(x)}\left({1\over 2a}\right)^2\sum^\dagger_\rho\gamma_\rho
\langle\psi_L(x_1)\bar\Psi_R(x+\rho) A_\nu^a(y)\rangle.
\label{ra3}
\end{eqnarray}
We make the Fourier transform in both sides of above recursion relations, 
obtain ($p,p'\not=0$)
\begin{eqnarray}
S_{LL}(p) \Lambda^a_{\mu LL}(p,p')S_{LL}(p')
&=&{i\over aM(p')}
S_{LL}(p) \Lambda^a_{\mu LR}(p,p')S_{RR}(p')
\sum_\rho\sin p'_\rho\gamma^\rho
\label{ra1p}\\
S_{LL}(p) \Lambda^a_{\mu LL}(p,p')S_{LL}(p')
&=&{i\over aM(p)}\sum_\rho\sin p_\rho\gamma^\rho
S_{RR}(p) \Lambda^a_{\mu RL}(p,p')S_{LL}(p')
\label{ra1p'}\\
S_{RR}(p) \Lambda^a_{\mu RR}(p,p')S_{RR}(p')
&=&{i\over aM(p')}
\sum_\rho\sin p'_\rho\gamma^\rho S_{LL}(p)\Lambda^a_{\mu LR}(p,p')S_{RR}(p').
\label{ra3p}
\end{eqnarray}
In these equations, the propagator of gauge boson $G^{ab}_{\nu\mu}(q)$ is
eliminated from the both sides of equations. 

Using these recursion relations (\ref{ra1p}-\ref{ra3p}), $S_{LL}(p)$ and
$S_{RR}(p)$ in eqs.(\ref{sll21},\ref{smm21}), we can compute the
vertex functions $\Lambda^a_{\mu RL}(p,p')$, $\Lambda^a_{\mu LR}(p,p')$ and
$\Lambda^a_{\mu RR}(p,p')$ in terms of the vertex function
$\Lambda^a_{\mu LL}(p,p')$ (\ref{normal}) that is obtained from perturbative
calculations in powers of the small gauge coupling.
\begin{eqnarray}
M(p')\Lambda^a_{\mu LL}(p,p')
&=&\Lambda^a_{\mu LR}(p,p')\left({i\over a}\right)
\sum_\rho\sin p'_\rho\gamma^\rho,
\label{v1p}\\
M(p)\Lambda^a_{\mu LL}(p,p')
&=&\left({i\over a}\right)\sum_\rho\sin p_\rho\gamma^\rho
\Lambda^a_{\mu RL}(p,p'),
\label{v1p'}\\
M(p')\Lambda^a_{\mu RR}(p,p')
&=&\left({i\over a}\right)
\sum_\rho\sin p'_\rho\gamma^\rho \Lambda^a_{\mu LR}(p,p').
\label{v3p}
\end{eqnarray}
Taking $\Lambda^a_{\mu LL}(p,p')$ to be eq.(\ref{normal}) at the leading 
order, we obtain
\begin{eqnarray}
\Lambda_{\mu RR}^{(1)}(p,p') &=& ig\left(\tau^a\over 2\right)\cos\left({p+p'\over 2}
\right)_\mu\gamma_\mu P_R,\label{vr}\\
\Lambda_{\mu LR}^{(1)}(p,p')\left({i\over a}\right)
\sin p'_\mu &=& {1\over2}M(p')ig\left(\tau^a\over 2\right)
\cos\left({p+p'\over 2}
\right)_\mu ,\label{vlr}\\
\left({i\over a}\right)
\sin p_\mu\Lambda_{\mu RL}^{(1)}(p,p') &=& {1\over2}M(p)ig\left(\tau^a\over 2\right)
\cos\left({p+p'\over 2}
\right)_\mu .
\label{vrl}
\end{eqnarray}
Thus, the coupling (\ref{3points}) between the gauge field and Dirac fermion (\ref{sdp})
is given by
\begin{equation}
\Lambda_{\mu c}^{(1)}=\Lambda_{\mu LL}^{(1)}+\Lambda_{\mu LR}^{(1)}
+\Lambda_{\mu RL}^{(1)}+\Lambda_{\mu RR}^{(1)}
\label{vdirac}
\end{equation}
These calculations can be straightforwardly generalized to higher orders of
the perturbative expansion in powers of the gauge coupling.
One can check that these results precisely obey the following Ward identity of
the exact $SU_L(2)$ chiral gauge symmetry $p', p\not= 0$
\begin{equation}
\left({i\over a}\right)(
\sin p_\mu-\sin p'_\mu)\Lambda_{\mu c}^{(1)}(p,p')=S_c^{-1}(p)-S_c^{-1}(p').
\label{gward}
\end{equation}
where the gauge coupling $g$ and generator ${\tau_a\over 2}$ are eliminated
from the vertex $\Lambda_{\mu c}$. These results are what we expected, since we are in the symmetric phase
($1\ll g_2<\infty$) where the exact $SU_L(2)$ chiral gauge symmetry is
realized by the vector-like spectrum excluding the low-energy states $p'\not=0$
and $p\not=0$. 

When $p', p\rightarrow 0$ and reach the threshold $\epsilon$ (\ref{pthre}) that
we discussed in section 5, the right-handed three-fermion state $\Psi_R^i(x)$
is supposed to disappear. The 1PI vertex functions $\Lambda_{\mu RR}$,
$\Lambda_{\mu RL}$, and $\Lambda_{\mu LR}$ relevant to $\Psi_R^i(x)$ vanish on
this threshold. The coupling vertex (\ref{vdirac}) between the gauge boson and
fermion turns out to be chiral in consistent with the $SU_L(2)$ chiral gauge
symmetry, ($p', p\rightarrow 0$) 
\begin{equation}
\left({i\over a}\right)(
\sin p_\mu-\sin p'_\mu)\Lambda_{\mu LL}^{(1)}(p,p')
=S_{L}^{-1}(p)-S_{L}^{-1}(p'),
\label{glward}
\end{equation}
where the propagator of chiral fermion $S_{L}^{-1}(p)$ is given by
eq.(\ref{sf}) and this Ward identity is realized by the chiral spectrum. Here
we stress again that the disappearance of the three-fermion (right-handed)
state is essential point to obtain continuum chiral gauge coupling in the
low-energy limit. However, we have to confess that similar to the threshold
(\ref{pthre}), eq.(\ref{glward}) is a plausible speculation for the time being,
since we need more evidence and computations to show whether or not it is ture.

The Ward identities (\ref{gward}) and (\ref{glward}) play an extremely
important role to guarantee that the gauge perturbation theory in the scaling
region ($1\ll g_2<\infty$) is gauge symmetric. To all orders of gauge coupling
perturbation theory, gauge boson masses vanish and the gauge boson propagator
is gauge-invariantly transverse. The gauge perturbation theory can be described
in the normal renormalization prescription as that of the QCD and QED theory.
In fact, due to the manifest $SU_L(2)$ chiral gauge symmetry and corresponding
Ward identities that are respected by the spectrum (vector-spectrum for
$p\not=0$ and chiral-spectrum for $p=0$) in this possible scaling regime, we
should then apply the Rome approach \cite{rome} (which is based on the
conventional wisdom of quantum field theories) to perturbation theory in the
small gauge coupling. It is expected that the Rome approach would work in the
same way but all gauge-variant counterterms are prohibited. 

\vskip1cm
\noindent
{\bf 7. The vacuum functional and gauge anomaly}
\vskip0.7cm

Provided the scenario of the gauge coupling and spectrum given in above
section, one should expect that the gauge field should not only chirally couple
to the massless chiral fermion of the $\psi_L^i$ in the low-energy regime, but
also vectorially couple to the massive doublers of Dirac fermion $\Psi_c^i$ in
the high-energy regime. In this section, we discuss the gauge anomaly and the
renormalization of gauge perturbation theory. 

We consider the following $n$-point 1PI functional:
\begin{equation}
\Gamma^{(n)}_{\{\mu\}}=
{\delta^{(n)}\Gamma(A')\over\delta A'_{\mu_1}(x_1)\cdot\cdot\cdot\delta 
A'_{\mu_j}
(x_j)\cdot\cdot\cdot\delta A'_{\mu_n}(x_n)},
\label{fun}
\end{equation}
where $j=1\cdot\cdot\cdot n, (n\geq 2)$ and $\Gamma(A')$ is the vacuum
functional. The perturbative computation of the 1PI vertex functions
$\Gamma^{(n)}_{\{\mu\}}$ can be straightforwardly performed by adopting the
method presented in ref.\cite{smit82} for lattice QCD. Dividing the integration
of internal momenta into 16 hypercubes where 16 modes live, we have 16
contributions to the truncated vertex functions. The region where is the chiral
fermion modes of continuum limit is defined as
\begin{equation}
\Omega=[0,\epsilon]^4,\hskip0.5cm p<\epsilon\ll {\pi\over 2},\hskip0.5cm
p\rightarrow 0 
\label{continuum}
\end{equation}
where the $\epsilon$ is the energy-threshold given by (\ref{pthre}),
on which $\Psi_R(x)$ disappears, in section 5.

As a first example, we deal with the vacuum polarization
\begin{equation}
\Pi_{\mu\nu}(p)=\sum_{i=1}^{16}\Pi^i_{\mu\nu}(p),\hskip0.5cm
\Pi^d_{\mu\nu}(p)=\sum_{i=2}^{16}\Pi^i_{\mu\nu}(p).
\label{vacuum}
\end{equation}
For the contributions
$\Pi^d_{\mu\nu}(p)$ from the 15 doublers $(i=2,...,16)$, we make a Taylor
expansion in terms of external physical momenta $p=\tilde p$ and the following
equation is {\it mutatis mutandis} valid 
\cite{smit82},
\begin{eqnarray}
\Pi^d_{\mu\nu}(p)&=&\Pi^\circ_{\mu\nu}(0)+\Pi^{d(2)}_{\mu\nu}(p)(\delta_{\mu\nu}
p^2-p_\mu p_\nu)\nonumber\\
&+&\sum^{16}_{i=2}\left(1-p_\rho|_\circ\partial_\rho-{1\over2}
 p_\rho p_\sigma|_\circ\partial_\rho\partial_\sigma\right)
\Pi^{con}_{\mu\nu}(p,m_i),
\label{smit}
\end{eqnarray}
where $|_\circ f(p)=f(0)$ and $m^i$ are doubler masses. The first and second terms are specific for the
lattice regularization. Since the 15 doublers are gauged as an $SU(2)$ QCD-like
gauge theory with propagator (\ref{sdp}) and interacting vertex (\ref{vdirac}), the
Ward identities (\ref{gward}) associated with this vectorial gauge symmetry result in the
vanishing of the first divergent term $\Pi^\circ_{\mu\nu}(0)$ and the gauge
invariance of the second finite contact term in eq.(\ref{smit}). We recall
that in Roma approach, this was achieved by enforcing Ward identities and 
gauge-variant counterterms. The third term
in eq.(\ref{smit}) corresponds to the relativistic contribution of the 15
doublers. The $\Pi^{con}_{\mu\nu}(p,m_i)$ is logarithmicly divergent and
evaluated in some continuum regularization. For doubler masses $m_i$ of
$O(a^{-1})$, the third term in eq.(\ref{smit}) is just finite and
gauge-invariant contributions. 

We turn to the contribution $\Pi^n_{\mu\nu}(p)$ of the massless
chiral mode that is in the first hypercube $\Omega=[-\epsilon,\epsilon]^4$
(\ref{continuum}).
We can use some regularization to calculate this 
contribution, 
\begin{equation}
\Pi^n_{\mu\nu}(p)=\Pi^{n(2)}_{\mu\nu}(p)(\delta_{\mu\nu}
p^2-p_\mu p_\nu).
\label{nsmit}
\end{equation}
The spectrum eq.(\ref{sf}) and gauge-coupling vertex eq.(\ref{normal}) with 
respect to
the chiral mode is $SU_L(2)$ chiral-gauge symmetric. The Ward identity
associated with this chiral gauge symmetry render eq.(\ref{nsmit}) to be gauge
invariant. The $\epsilon$-dependence ($\ln\epsilon$) in eq.(\ref{nsmit}) has to be exactly
cancelled
out from those contributions (\ref{smit}) from doublers, because 
the continuity of 1PI vertex functions in momentum 
space. 
In summary, the total vacuum polarization $\Pi_{\mu\nu}(p)$ contains two parts:
(i) the vacuum polarization of the chiral mode $\psi^i_L$ in some continuum 
regularization; (ii) gauge invariant finite terms stemming from doublers'
contributions. The second part is the same as the perturbative lattice 
QCD, and can be subtracted away in normal renormalization prescription.

The second example is the 1PI vertex functions $\Gamma^{(n)}_{\{\mu\}}(\{p\}) (n\geq
4)$,
\begin{eqnarray}
\Gamma^{(n)}_{\{\mu\}}(\{p\})&=&\sum^{16}_{i=1}
\Gamma^{(n)i}_{\{\mu\}}(\{p\},m_i)\hskip0.5cm n\geq 4,
\label{n=4}\\
\{p\}&=&p_1,p_2,\cdot\cdot\cdot\nonumber\\
\{\mu\}&=&\mu_1,\mu_2,\cdot\cdot\cdot,\nonumber
\end{eqnarray}
where internal momentum integral is analogously divided into the contributions
from sixteen sub-regions of the Brillouin zone where sixteen modes live. Based on gauge invariance
and power counting, one concludes that up to some gauge invariant finite terms,
the $\Gamma^{(n)}_{\{\mu\}}(\{p\}) (n\geq 4)$ (\ref{n=4}) contain the 15
continuum expressions for 15 massive ($m_i$) Dirac doublers and one for the
massless Weyl mode. The 15 doubler contributions vanish for $m_i\sim
O(a^{-1})$. The $n$-point 1PI vertex functions (\ref{n=4}) end up with their
continuum counterpart for the Weyl fermion and some gauge invariant finite
terms. These finite gauge invariant terms come from  doublers' contributions
are similar to those in the lattice QCD, and can be subtracted away in the
normal renormalization prescription. 

The most important contribution to the vacuum functional is the triangle graph
$\Gamma_{\mu\nu\alpha} (p,q)$, which is linearly divergent. Again, dividing the
integration of internal momenta into 16 hypercubes, one obtains\cite{smit82}
\begin{eqnarray}
\Gamma_{\mu\nu\alpha}(p,q)&=&\sum^{16}_{i=1}\Gamma^i_{\mu\nu\alpha}(p,q)
\nonumber\\
\Gamma^i_{\mu\nu\alpha}(p,q)&=&\Gamma^{i(\circ)}_{\mu\nu\alpha}(0)+
p_\rho\Gamma^{i(1)}_{\mu\nu\alpha ,\rho}(0)
+q_\rho\Gamma^{i(1)}_{\mu\nu\alpha ,\rho}(0)\nonumber\\
&+&\left(1-|_\circ - p_\rho|_\circ\partial_\rho - q_\rho|_\circ\partial_\rho
\right)
\Gamma^{con}_{\mu\nu\alpha}(p,q,m_i),
\label{smit1}
\end{eqnarray}
where $\Gamma^{con}_{\mu\nu\alpha}(p,q,m_i)$ is evaluated in some continuum
regularization. As for the 15 contributions of Dirac doublers
($i=2\cdot\cdot\cdot 15$), the first three terms in eq.(\ref{smit1}) 
are zero for the vector-like
Ward identity (\ref{gward}). The non-vanishing contributions are the same as the
15 copies of the $SU(2)$ vectorial gauge theory of massive Dirac fermions.
These contributions are gauge-invariant and finite (as $m_i\sim O(a^{-1})$),
thus, disassociate from the gauge anomaly. 

The non-trivial contribution of the
chiral mode in the hypercube $\Omega =[-\epsilon,\epsilon]^4$ is given by
\begin{eqnarray}
\Gamma^{i=1}_{\mu\nu\alpha}(p,q)\!&\!=\!&\!\int_\Omega\!
{d^4k\over(2\pi)^4}\tr\!\left[S(k\!+\!{p\over2})\Gamma_\mu(k)S(k\!-\!{p\over2})
\Gamma_\nu(k\!-\!{p\!+\!q\over2})S(k\!-\!{p\over2}\!-\!q)\Gamma_\alpha(k\!-\!
{q\over2})\right]\nonumber\\
&&+(\nu\leftrightarrow\alpha),
\label{tri}
\end{eqnarray}
where the propagator $S(p)$ and vertex $\Gamma_\mu$ are given by
eqs.(\ref{sf},\ref{normal}). Other contributions containing anomalous vertices
$(\psi\bar\psi AA, \psi\bar\psi AAA)$ vanish within 
the hypercube $\Omega =[-\epsilon,\epsilon]^4$. As well known, eq.~(\ref{tri}) 
is not gauge invariant. To evaluate eq.(\ref{tri}),
one can use some continuum regularizations. As a result, modulo possible local
counterterms, we obtain the consistent gauge anomaly for the non-abelian 
chiral gauge theories as the continuum one: 
\begin{equation}
\delta_g\Gamma(A')=-{ig^2\over24\pi^2}\int d^4x
\epsilon^{\alpha\beta\mu\nu}\tr\theta_a(x)\tau_a\partial_\nu 
\left[A_\alpha(x)\left(\partial_\beta A_\mu+{ig\over2}A_\beta (x)A_\mu(x)
\right)\right],
\label{anomaly}
\end{equation}
where the gauge field $A_\mu={\tau^a\over2} A^a_\mu$. 
The $SU_L(2)$ chiral gauge theory is anomaly-free for
$\tr(\tau^a,\{\tau^b,\tau^c\})=0$, and the gauge current
\begin{equation}
J^a_\mu=i\bar\psi_L\gamma^\mu{\tau^a\over2}\psi_L={\delta\Gamma(A)\over \delta
A_\mu^a(x)}\hskip0.5cm \partial^\mu J^a_\mu=0
\label{conser}
\end{equation}
is covariantly conserved and gauge invariant. In the following paragraph, we
discuss how we achieve the correct gauge anomaly (\ref{anomaly}) from the gauge
symmetric action (\ref{action}).

A most subtle property of the naive lattice chiral gauge theory is the
appearance of 16 modes. Each mode produces the chiral gauge anomaly with
definite axial charge $Q_5$\cite{smit82}, such that the finite (regularized)
theory is anomaly-free and the chiral gauge symmetry is perfectly preserved. As
has been seen, the 15 doublers decoupled as massive Dirac fermions that are
vector-gauge-symmetric (\ref{gward}). Thus, they decouple from the gauge anomaly as
well. Only the anomaly associated with the normal (chiral) mode of the $\psi^i_L$ is
left and is the same as the continuum one. The condition is the disappearance
of the right-handed three-fermion state $\Psi^i_R$ in the low-energy limit.
It seems surprising that we start from
a gauge symmetric action and we end up with the correct gauge anomaly.
Normally, one may claim that the anomaly has to come from the explicit breaking
of the chiral gauge symmetry in a regularized action (e.g., a Wilson term). This
statement is indeed correct if regularized actions are bilinear in fermion fields,
since this is nothing but what the ``no-go" theorem asserts. However, we run
into the dilemma that the gauge anomaly is independent of any explicitly
breaking parameters (e.g., the Wilson parameter $r$ and fermion masses). In
fact, the most essential and intrinsic {\it raison d'\^etre} of producing the
correct gauge anomaly is ``decoupling doublers'' rather than ``explicitly
breaking of chiral gauge symmetries ''. If we adopt a bilinear action to decouple
doublers, we must explicitly break chiral gauge symmetry, which is just a
superficial artifact in bilinear actions. However, if we give up the bilinearity of regularized
actions in fermion fields and find a chiral-gauge-invariant way to decouple
doublers, we should not surprise to achieve the correct gauge anomaly 
(\ref{anomaly}). These discussions on the gauge anomaly is not complete yet. It
is related again to the possibility of filling three-fermion states 
(\ref{emom}) into the Dirac sea, further discussions are necessary. Reads are
suggested to the papers by Nielsen and Ninomiya \cite{nn91} and Creutz 
\cite{creutz}.

\vskip1cm
\noindent
{\bf 8. The anomalous global current}
\vskip0.7cm

Non-conservation of fermion numbers is an important feature of the Standard
Model. A successful regularization of chiral gauge theories should give this
feature in the continuum limit \cite{ep,b92,over}. In EP's approach of 
multifermion couplings, inspired by the origination of the axial anomaly in
lattice QCD, it is suggested that the anomalous global current should be
originated from the explicit breaking of the global symmetry at the tree
level. 

In this section, we show a possibility that this global anomaly
can be consistently obtained from the explicit symmetric action (\ref{action})
with the multifermion coupling, if the composite right-handed fermion 
$\Psi^i_R(x)$ disappears into three chiral constituents in the low-energy
limit, and the theory is free from local gauge symmetry breaking and 
the gauge anomaly. 

Our action (\ref{action}) processes the $U_L(1)$ and $U_R(1)$ global chiral 
symmetries. At tree level, the theory is invariant under the following 
transformations:  
\begin{equation}
\psi_L^i\rightarrow e^{i\theta_L} \psi_L^i\hskip1cm
\psi_R\rightarrow e^{i\theta_R}\psi_R.
\label{dd}
\end{equation}
These global symmetries lead to the conservation
of the singlet chiral fermion currents,
\begin{eqnarray}
\partial_\mu j^\mu_L(x)&=&0,\hskip0.5cm
j^\mu_L=i\bar\psi_L^i\gamma^\mu\psi^i_L\label{lcu}\\
\partial_\mu j^\mu_R(x)&=&0,\hskip0.5cm
j^\mu_R=i\bar\psi_R\gamma^\mu\psi_R +O(a^2),
\label{rcu}
\end{eqnarray}
which are Noether currents. Eqs.(\ref{lcu},\ref{rcu}) correspond to the conservation
of fermion numbers. However, as we know, eq.(\ref{lcu}) should be anomalous. 

In order to see whether the conservation of the currents is violated when the
chiral gauge field is coupled to chiral fermions, we consider the source
currents $\langle j^\mu_L(x)\rangle$ and $\langle j^\mu_R(x)\rangle$ defined as
\begin{eqnarray}
\langle j^\mu_{L}(x)\rangle &=&{\delta \Gamma\over\delta 
V^{L}_\mu(x)};
\hskip1cm \delta V^{L}_\mu(x)= -\partial_\mu\theta_{L}(x);\label{sourcel}\\
\langle j^\mu_{R}(x)\rangle &=&{\delta \Gamma\over\delta 
V^{R}_\mu(x)};
\hskip1cm \delta V^{R}_\mu(x)= -\partial_\mu\theta_{R}(x).
\label{sourcer}
\end{eqnarray}
Under the variations $\delta_L$ and $\delta_R$ of these $U(1)$-phases
$\theta_{L}(x)$ and $\theta_{R}(x)$,
the effective action $\Gamma$ is transformed  (up to $O(\theta_{L})$ and
$O(\theta_{R})$)
\begin{eqnarray}
\delta_{L}\Gamma &=&\int d^4x\delta V^{L}_\mu(x)\langle j^\mu_{L}(x)
\rangle=\int d^4x \theta_{L}(x)\partial_\mu
\langle j^\mu_{L}(x)\rangle,\label{deltall}\nonumber\\
\delta_{R}\Gamma &=&\int d^4x\delta V^{R}_\mu(x)\langle j^\mu_{R}(x)
\rangle=\int d^4x \theta_{R}(x)\partial_\mu
\langle j^\mu_{R}(x)\rangle,
\label{deltarr}
\end{eqnarray}
where
\begin{eqnarray}
\delta_{R}\Gamma &=& \Gamma(A_\mu+\delta V^{R}_\mu(x))-\Gamma(A_\mu),
\label{variationr}\\
\delta_{L}\Gamma &=& \Gamma(A_\mu+\delta V^{L}_\mu(x))-\Gamma(A_\mu).
\label{variationl}
\end{eqnarray}
In our action (\ref{action}), the $\psi_R$ does not couple to the chiral
gauge field and this decoupling strictly holds due to the Ward identity
(\ref{wa}). Thus, the $\langle j^\mu_R(x)\rangle$ defined formally in
eq.(\ref{sourcer}) is gauge invariant 
\begin{equation}
\delta_g\langle j^\mu_R(x)\rangle=0,
\label{1}
\end{equation}
and eq.(\ref{variationr}) becomes,
\begin{equation}
\delta_R\Gamma(A)=0.
\label{gs}
\end{equation}
This leads to the conservation of the right-handed current, 
\begin{equation}
\partial_\mu\langle j^\mu_R(x) \rangle=0.
\label{2}
\end{equation} 
We find that the Ward identity of the $\psi_R$-shift-symmetry and decoupling between
the gauge field and the spectator fermion $\psi_R(x)$ are crucial to the conservation
of the right-handed fermion numbers (\ref{2}).

We have shown in section 7, under a gauge transformation $\delta_g$,
\begin{equation}
\delta_g\Gamma_{\mu\nu\alpha}(p,q)\not=0,\hskip1cm
\delta_g\Gamma^{(n)}_{\{\mu\}}=0\hskip0.5cm (n\not=3).
\label{nm}
\end{equation}
The vacuum functional $\Gamma$ is just the same as the continuum counterpart 
up to some gauge-invariant finite terms. In the anomaly-free $SU_L(2)$ case
($\delta_g\Gamma=0$), one may conclude that the current $\langle j^\mu_L(x)
\rangle$ defined in eq.(\ref{sourcel}) is gauge invariant. 
\begin{equation}
\delta_g\langle j^\mu_L(x)
\rangle=\delta_g{\delta \Gamma\over\delta 
V^L_\mu(x)}={\delta\delta_g\Gamma \over\delta V^L_\mu(x)}=0.
\label{gauin}
\end{equation}
However, this is not true. The order of the differentiations $\delta_g$ and 
$\delta_L$ can not be exchanged. We know that in our action (\ref{action}), the
left-handed variation
\begin{equation}
\delta V^L_\mu(x)= - \partial_\mu\theta_L(x)
\label{vl}
\end{equation}
can be considered as a commuting $U_L(1)$ factor in the $SU_L(2)$ chiral 
gauge group, i.e.
\begin{equation}
\tilde A_\mu=A_\mu+V_\mu^L,\hskip1cm (A_\mu={\tau^a\over2} A^a_\mu).
\label{va}
\end{equation}
Actually,
this is a $SU_L(2)\otimes U_L(1)$ chiral gauge group and there is a mixing
anomaly\cite{preskill91}, 
\begin{eqnarray}
\delta_L\Gamma &=& C_1{ig^2\over32\pi^2}\int d^4x
\theta_L\tr\left(F_{\mu\nu}\tilde F^{\mu\nu}\right),\label{arbi1}\\
\delta_g\Gamma&=& C_2{ig\over16\pi^2}\int d^4x
\tilde F^{\mu\nu}_1\tr\left(\theta_g\partial_\mu A_\nu\right),
\label{arbi}
\end{eqnarray}
where $\theta_g=\theta^a_g\tau_a$ is the $SU_L(2)$ transformation parameter and 
\begin{equation}
F^{\mu\nu}_1=\partial^\mu V^\nu-\partial^\nu V^\mu .
\label{f1}
\end{equation}
The reason is that one of the Pauli
matrices ${\tau^a\over2}$ in the triangle graph is replaced by the generator
(identity) of the $U_L(1)$, i.e., the $U_L(1)$ global current, therefore 
the vanishing of the $SU_L(2)$ anomaly for
$\tr(\tau^a,\{\tau^b,\tau^c\})=0$ is no longer true. Note that in 
eqs.(\ref{arbi1},\ref{arbi}), we only consider
the triangle diagram $(n=3)$, since 
\begin{equation}
\delta_L\Gamma^{(n)}_{\{\mu\}}=0,\hskip0.3cm\delta_g\Gamma^{(n)}_{\{\mu\}}\hskip0.3cm 
(n\not=3)
\label{3}
\end{equation}
for being gauge-invariant, as we discussed
in section 7. 

The mixing anomaly (\ref{arbi}) has arbitrariness $C_1, C_2$ $(C_1+C_2=1)$,
which arise because the triangle graphs with one insertion of the $U_L(1)$
global current determine the $\Gamma_{\mu\nu\alpha}(A')$ up to a local
counterterm. As the Feynman diagrams determine the effective action $\Gamma
(A)$ only up to an arbitrary choice of local counterterms, we are allowed to
add local counterterms into the effective action 
\begin{equation}
\Gamma'(A')=\Gamma(A')+\Gamma_{c.t.}(A'),
\label{reg}
\end{equation}
which is equivalent to the
re-definition of the chiral fermion current eq.(\ref{sourcel}). Due to the fact
that the effective action $\Gamma(A')$ we obtained for the $SU_L(2)$ case is free from local
gauge-symmetry-breaking terms and the non-local gauge anomaly, the
arbitrariness in eq.(\ref{arbi}) can be fixed 
\begin{equation}
C_1=1,\hskip0.5cm C_2=0
\label{c1c2}
\end{equation}
by choosing an
adequate local counterterm. As a result, the effective action and the
left-handed current are gauge invariant,
\begin{equation}
\delta_g\Gamma'(A)=0,\hskip1cm \delta_g
\langle j^{'\mu}_L\rangle=0,
\label{pp}
\end{equation}
where $\langle j^{'\mu}_L\rangle$ is  the re-definition of the left-handed
chiral fermion current eq.(\ref{sourcel}). From eq.(\ref{deltall}), one obtains
\begin{equation}
\delta_L\Gamma'={ig^2\over32\pi^2}\int d^4x\theta_L\tr\left(F_{\mu\nu}\tilde 
F^{\mu\nu}\right);\hskip0.5cm\partial_\mu \langle j^{'\mu}_L\rangle =
{ig^2\over32\pi^2}\tr\left(F_{\mu\nu}\tilde F^{\mu\nu}\right).
\label{fer}
\end{equation}
This is just the desired result, which shows the left-handed fermion number is
violated by the $SU(2)$ instanton effect. 

We stress again that the exact chiral gauge symmetry of a regularized action,
gauge invariant 1PI vertex functions (absence of local gauge variant terms and
anomaly-free, e.g., the $SU_L(2)$ case and Standard Model) and the right-handed
composite fermion $\Psi^i_R(x)$ dissolving into three free chiral fermions plays
an extremely crucial role in obtaining the gauge-invariant
chiral-fermion-current (\ref{pp}) and its non-conservation (\ref{fer}). The reason for obtaining the
correct anomaly (\ref{fer}) coincides with that considered in the approach of
\cite{th94}. 

One may ask why this global anomaly comes from an explicit
$U_L(1)$ symmetric action (\ref{action}). This question is raised because of our
knowledge of the lattice QCD where the axial current anomaly, 
\begin{equation}
\partial_\mu j^\mu_5=
{ig^2\over16\pi^2}\tr\left(F_{\mu\nu}\tilde F^{\mu\nu}\right)
\label{aa}
\end{equation}
is due to the flavour $SU_L(3)\otimes SU_R(3)$ asymmetric Wilson term. However,
{\it a priori}, we have no dynamical reason to expect that the non-conservation
of fermion number is due to the explicit breaking of the $U_L(1)$ symmetry at
the cutoff level. A non-gauge-invariantly regularized action with the
$U(1)$-asymmetry, (e.g., Majorana Wilson-Yukawa coupling\cite{rome2}) does not
lead to the correct anomaly eq.(\ref{fer})\cite{aoki92}, unless we force the
Ward identities of the chiral gauge symmetry to be obeyed by tuning
counterterms \cite{rome2}. It is still unknown whether chiral gauge symmetric
and the U(1) asymmetric model such as $SU(5)$\cite{ep} and $SO(10)$\cite{gpr}
give the correct anomaly to the conservation of chiral fermion numbers. It is
expected\cite{ns92,nn91} that the correct anomaly is unlikely to be produced in
these models because explicit-breaking contributions to $\partial_\mu j^\mu_L$
will normally have a quite different $\psi$-field-dependence. In fact, the
anomaly (\ref{fer}) disappears as the gauge field turns off, the conservation
of the global current must be related to the explicit $U_L(1)$ and $U_R(1)$
symmetries of the action (\ref{action}). These discussions are not completely
clarified and they need further studies. To look at this problem, it is helpful
to read the papers by Nielsen and Ninomiya \cite{nn91} and Creutz \cite{creutz}
about their intuitive understanding of anomalies. 

\vskip1cm
\noindent
{\bf 9. Summary}
\vskip0.7cm

In summary, we present a possible lattice chiral gauge theory with the large
multifermion coupling. We discuss how to realize EP's idea in this model and
study relevant problems of regularizing chiral gauge theories, which are listed
in the introduction section. We advocate the segment ($1\ll g_2<\infty$) to 
be the scaling region for defining a continuum chiral gauge theory. In the light of
studies presented in this paper, it seems to us that a certain multifermion
coupling model with EP's idea has chances to work. However, in the multifermion
coupling model proposed in this paper, there are many things such as no NJL
gauge symmetry breaking, decoupling of doublers, disappearance of the
right-handed composite fermion $\Psi^i_R$, etc.~that need to be further
clarified and demonstrated. For the controversy of the issue, it is suggested
to study multifermion coupling models and solve relevant problems in 1+1
dimensions \cite{en}. It is worthwhile to mention that applying EP's idea only
to mirror fermions residing in one wall of Kaplan's model \cite{kaplan} is a
proposal that may work \cite{cx}. 
 
In conclusion, we advocate that the further study of EP's idea and multifermion
couplings for regularizing chiral gauge theories be necessary. 
I thank Profs.~G.~Preparata, M.~Creutz, H.B.~Nielsen and E.~Eichten
for discussions.


\begin{thebibliography}{99}

\bibitem{nn81}
H.B.~Nielsen and M.~Ninomiya, {\sl Nucl.~Phys.} {\bf B185} (1981) 20, {\it
ibid.} {\bf B193} (1981) 173, {\sl Phys.~Lett.} {\bf B105} (1981) 219.

\bibitem{ep}
E.~Eichten and J.~Preskill, {\sl Nucl.~ Phys.} {\bf B268} (1986) 179.

\bibitem{xue}
G.~Preparata and S.-S.~Xue, {\sl Phys.~Lett.} {\bf B264} (1991) 35;
{\it ibid} {\bf B335} (1994) 192; {\bf B329} (1994) 87;
{\bf B325} (1994) 161; {\bf B377} (1996) 124;
{\sl Nucl.~Phys.} {\bf B26} (Proc.~Suppl.) (1992) 501;
{\sl Nucl.~Phys.} {\bf B30} (Proc.~Suppl.) (1993) 647.\\
S.-S.~Xue {\sl Nucl.~Phys.} (Proc.~Suppl.) {\bf B47} (1996) 583.

\bibitem{ns92}
H.B.~Nielsen and S.E.~Rugh, {\sl Nucl.~ Phys.} {\bf 29B} (Proc.~Suppl.) 
(1992) 200.

\bibitem{b92} T.~Banks, {\sl Phys.~Lett.} {\bf B272}
(1991) 75; T.~Bank and A.~Dabholkar,
{\sl Nucl.~ Phys.} {\bf 29B} (Proc.~Suppl.) 
(1992) 46.

\bibitem{njl}
Y.~Nambu and G.~Jona-Lasinio, {\sl Phys. Rev.} {\bf 122} (1961) 345.

\bibitem{gpr}
M.F.L.~Golterman, D.N.~Petcher and E.~Rivas, {\sl Nucl.\ Phys.} {\bf B395} 
(1993) 597.

\bibitem{ss}
J.~Smit, {\sl Acta Physica Polonica} {\bf B17} (1986) 531;\\
P.D.V.~Swift, {\sl Phys.~Lett.} {\bf B145} (1984) 256.

\bibitem{p}
D.N.~Petcher, {\sl Nucl.~Phys.}(Proc.~Suppl.) {\bf B30} 
(1993) 52, references there in.

\bibitem{ys93}
Y.~Shamir, {\sl Phys.~Rev.~Lett.} {\bf 71} (1993) 2691;
{\sl Nucl.~Phys.}(Proc.~Suppl.) {\bf B47} (1996) 212.

\bibitem{nn91}
H.B.~Nielsen and M.~Ninomiya, {\sl Int.~J.~of Mod.~Phys.} {\bf A6} (1991) 2913;
H.B.~Nielsen and S.E.~Rugh, {\sl Nucl.~ Phys.} {\bf 29B} (Proc.~Suppl.) 
(1992) 200.

\bibitem{gp}
M.F.L.~Golterman, D.N.~Petcher, {\sl Phys.~Lett.} {\bf B225} 
(1989) 159.

\bibitem{wilson}
K.~Wilson, in {\it New phenomena in subnuclear physics\/} 
(Erice, 1975) 
ed.\ A.~Zichichi (Plenum, New York, 1977).

\bibitem{xue95} 
S.-S.~Xue, {\sl Phys.~Lett.} {\bf B381} (1996) 277 and hep-lat/9605005.

\bibitem{mont} 	I.~Montvay, {\sl Nucl. Phys.} {\bf B29} (Proc.~Suppl.)
(1992) 159, references therein.

\bibitem{rome} 
A.~Borrelli, L.~Maiani, G.C.~Rossi, R.~Sisto and M. Testa, {\sl
Nucl.~ Phys.} {\bf B333} (1990) 335; {\sl Phys.~Lett.} {\bf B221} (1989) 360;\\
L.~Maiani, G.C.~Rossi, and M. Testa, {\sl Phys.~Lett.} {\bf B261} (1991) 479;
{\it ibid} {\bf B292} (1992) 397;\\ 
L.~Maiani, {\sl Nucl.~ Phys.} (Proc.Suppl.) {\bf B29} (1992) 33. 

\bibitem{smit82}
L.H.~Karsren and J.~Smit, {\sl Nucl.~ Phys.} {\bf B183} (1981) 103.

\bibitem{creutz}
M.~Creutz, {\sl Nucl.~Phys.}(Proc.~Suppl.)  {\bf B42} (1995) 56,
{\sl Phys.~Rev.} {\bf D52} (1995) 2951;\\
M.~Creutz and I.~Horv\'ath,
{\sl Phys.~Rev.} {\bf D50} (1994) 2297;

\bibitem{over}
R.~Narayanan and H.~Neuberger, {\sl Nucl.~ Phys.} {\bf B443} (1995) 305, 
references there in.

\bibitem{preskill91}
J.~Preskill, {\sl Ann.~Phys.} {\bf 210} (1991) 323.

\bibitem{th94}
G.~'t Hooft, {\sl Phys.~Lett.} {\bf B349} (1995) 491;
P.~Hern\'andez and R.~Sundrum, {\sl Nucl.~ Phys.} {\bf B455} (1995) 287.

\bibitem{rome2}
L.~Maiani, G.C.~Rossi and M.~Testa, {\sl Phys.~Lett.} {\bf 292} (1992) 397.

\bibitem{aoki92}
S.~Aoki, {\sl Nucl.~ Phys.} {\bf 29B} (Proc.~Suppl.) (1992) 171.

\bibitem{en}
Private communications with H.B.~Nielsen and E.~Eichten.

\bibitem{cx}
M.~Creutz and S.-S.~Xue, in progress.

\bibitem{kaplan}
D.~B.~Kaplan, {\sl Phys.~ Lett.} {\bf B288} (1992) 342.

\end{thebibliography}
\end{document}